\documentclass{article}

\usepackage{graphicx} 
\usepackage{braket}
\usepackage{amsmath}
\usepackage{caption}
\usepackage{subcaption}
\usepackage{authblk}
\usepackage[backend=biber,style=numeric,sorting=none]{biblatex} 
\begin{document}

\title{Optical cavity characterization with a mode-matched heterodyne sensing scheme}

\author[1]{Aaron D. Spector \thanks{aaron.spector@desy.de} }
\author[1]{Todd Kozlowski}

\affil[1]{Deutsches Elektronen-Synchrotron DESY, Notkestr. 85, 22607 Hamburg, Germany}



\maketitle
\begin{abstract} 
We describe a technique for measuring the complex reflectivity of an optical cavity with a resonant local oscillator laser and an auxiliary probe laser, each coupled via opposite ends of the cavity. A heterodyne sensing scheme is then used to observe the phase and amplitude of the interference beat-note between the promptly reflected field and the cavity transmitted field injected through the far mirror. Since the local oscillator laser must pass through the cavity before interfering with the probe laser these measurements are not only independent of the spatial coupling of either laser to the cavity, but also obtained at the in-situ position of the cavity Eigenmode. This technique was demonstrated on a 19\,m cavity to measure the individual transmissivities of each of the mirrors as well as the round trip optical losses to an accuracy of several parts per million.
\end{abstract}


\section{Introduction}

High-finesse optical cavities are excellent tools for a variety of applications ranging from gravitational wave detection \cite{ligo2015,virgo2015,capocasa2018}, to searches for particles beyond the standard model \cite{hoogeveen1991,ehret2010new,bahre2013any,alps2022}, and vacuum magnetic birefringence \cite{della2016pvlas,cadene2014vacuum}, as well precision tests of quantum geometry at the Plank scale \cite{chou2017,vermeulen2021experiment}. While the precise characterization of the cavities used in these experiments is very important, it can prove particularly challenging to discriminate between cavity mirror transmissivity and excess losses. 

Although it is possible to interrogate cavity optic surface roughness and flatness ex-situ to project intra-cavity losses due to scattering \cite{straniero15}, these techniques are generally  dependent on the Eigenmode position on the mirrors and vulnerable to changes in the time between measurement and installation. Measurements of the cavity ring-down use the decay constant of the power of light in transmission and or reflection of a cavity to determine the total round-trip attenuation to high accuracy in-situ. Ring-down measurements, however, require a priori assumptions of one or both of the mirror transmissivities to determine round trip excess optical loss \cite{isogai13}. Locke et al. \cite{locke2009} used measurements of the reflection and transmission coefficients of an optical cavity to thoroughly characterize their resonator, including mirror transmission and losses. The measurement accuracy for these parameters, however, is substantially limited by uncertainties in photodiode power calibration. 


Using heterodyne interferometery, both the the phase and amplitude of a `probe field' can be measured in reflection of the cavity, relative to a second `local oscillator field' also incident on the cavity. By scanning the frequency of the probe field over the cavity resonance, its' complex reflectivity can be measured independent of the calibration of the photodetectors. Such a technique was performed by Slagmolen et al. \cite{slagmolen2000} to directly assess the linewidth, FSR, and coupling condition of a cavity. 

If both the local oscillator and probe field injected into the cavity at the same mirror as was done by Slagmolen et al., both the light in the fundamental mode of the cavity as well as the higher order modes of both lasers will interact and have an effect on the phase and amplitude of the interference beat-note. Calculating the cavity parameters from this measurement therefore requires precise knowledge of the spatial coupling of the lasers to the cavity.

In this article, we propose a technique which leverages the advantages of this scheme, while also making use of the spatial filtering of the cavity itself.
If the probe and local oscillator fields are instead injected via opposite ends of the cavity, only the light in the cavity Eigenmode will contribute to the interference beat-note. This technique thus requires no prior knowledge of the spatial coupling of either laser to the cavity, leading to an improvement of more than a factor of 10 in the accuracy of measurements when compared to Slagmolen et al. 

In the following, we show how the complex frequency dependent reflectivity can be examined to reveal several critical cavity parameters, a technique which has been previously applied to superconducting microwave cavity characterization \cite{probst2015,baity2024circle}. This method is then demonstrated on a 19\,m cavity to measure the individual mirror transmissivities and round-trip excess optical losses with an accuracy on the order of 2\,ppm.

\section{Theory}
Consider a Fabry-Perot cavity consisting of two mirrors, $M_1$ and $M_2$, with reflection and transmission coefficients $r_1$ and $t_1$, and $r_2$ and $t_2$, respectively. For each mirror the reflectivity $R$, transmissivity $T$, and excess optical losses $L$ in terms of power can be represented by the following equations,
\begin{equation}
    R_{1,2} = \left|r_{1,2}\right|^2
\end{equation}
\begin{equation}
    T_{1,2} = \left|t_{1,2}\right|^2
\end{equation}
Conservation of energy requires that, for each mirror,
\begin{equation}
    1 = R + T + L
\end{equation}
It should be noted here that by using the definition above and assuming the losses on the path between the cavity mirrors are negligible, the excess losses per reflection are included the mirrors coefficients of reflection.

For a monochromatic laser field $E_i$ coupled into the cavity via $M_1$, the field circulating in the cavity $E_{\rm cav}$ can be expressed as an infinite sum \cite{siegman86},
\begin{equation}
    E_{\rm cav} = \sum_{n=0}^\infty t_1 E_i  \left(r_1 r_2 e^{i kl}\right)^n
\end{equation}
Here, the term $r_1 r_2 e^{i 2kl}$ expresses the attenuation and phase shift experienced by the field in each round trip through the cavity due to its optical path length $l$ and the reflection coefficients of the mirrors. In this equation, $k$ is the wave vector of the input field and is equal to $2\pi/\lambda$ with $\lambda$ being wavelength of the input field.

Since the term in the summation is a geometric series, the previous expression is equivalent to the following equation,
\begin{equation}
    E_{\rm cav} = \frac{t_1 E_i}{1-r_1 r_2 e^{i kl}}
    \label{eq:cav}
\end{equation}
It is apparent from this equation that the cavity will be on resonance and achieve its maximum circulating field when $kl$ is some integer multiple of $2\pi$. This occurs when the round-trip optical path length of the cavity is some integer multiple of $\lambda$. 

\subsection{Complex Cavity Reflectivity}

In the case of perfect mode matching between the input field and the cavity, the reflected field can be represented as  the following superposition of the promptly reflected input field, given by $r_1 E_i$, and the fraction of the cavity circulating field that leaks through the input mirror, $t_1 r_2 e^{i kl} E_{\rm cav}$.  
\begin{equation}
    E_\mathrm{ref} = r_1 E_{\rm i} - t_1 r_2 e^{ikl} E_{\rm cav}
\end{equation}
Here we adopt the beamsplitter sign convention, wherein the transmitted leakage field gains a $\pi$ phase shift relative to the promptly reflected field. Using Eq. \ref{eq:cav} for $E_{\rm cav}$, $E_{\rm ref}$ then becomes
\begin{equation}
    E_{\rm ref} = E_i  \left( r_1  - \frac{t_1^2  r_2 e^{i kl}}{1-r_1 r_2 e^{i kl}} \right).
    \label{Eq:E_r raw}
\end{equation}
The frequency-dependent cavity complex reflectivity $\mathcal{R}$ is given by the ratio the the reflected and incident fields. For a high-finesse cavity near resonance, this can be approximated as
\begin{equation}
    \mathcal{R}(\Delta \nu) \equiv \frac{E_{\rm ref}}{E_i } \approx  1- \frac{T_1}{\frac{A}{2}-i\frac{\Delta\nu}{f_0}},
    \label{Eq:E_r}
\end{equation}
where $A = T_1 + L_1 + T_2 + L_2$ is total attenuation of the circulating power during a single cavity round-trip, $\Delta\nu$ is the frequency difference between the input laser field and the nearest cavity resonance, and $f_0$ is the cavity free spectral range.

\begin{figure}
    \centering
    \includegraphics[width=0.5\linewidth]{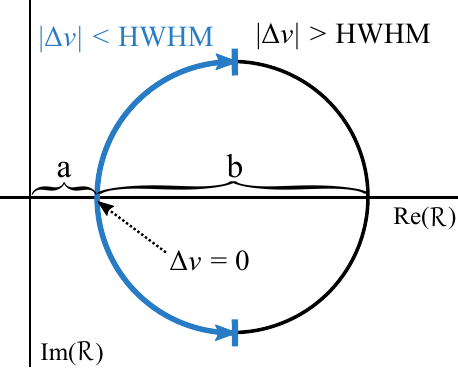}
    \caption{Illustration of the field reflected from the cavity plotted on the complex plane. The quantity $a+b$ gives the amplitude of the input field $E_i$. The point on the circle closest to the origin shows when the cavity resonance condition is met ($\Delta \nu = 0$), with $a$ giving the amplitude of the reflected field when the cavity is on resonance. As the frequency $\Delta \nu$ is varied, $\mathcal{R}(\Delta \nu)$ traces out a circle with diameter `b' which can be used to calculate the quantity $2T_1 / A$. 
    Here, the blue semicircle shows frequencies that lie within the cavity linewidth ($|\Delta \nu| <$ HWHM).}
    \label{fig:circle-cartoon}
\end{figure}

Figure \ref{fig:circle-cartoon} shows a representation of Eq. \ref{Eq:E_r raw} on the complex plane. As the frequency difference between the laser and the cavity resonance, $\Delta\nu$, is scanned, the equation traces out a circle on the complex plane revealing several important cavity parameters. The ratio of the circle's diameter to the distance from the origin of its furthest point, $\frac{b}{a+b}$ is equivalent to the ratio $2T_1/A$. For an under-coupled cavity, this ratio is less than one, as this diagram shows, and origin of the plot lies outside the circle. For a critically coupled cavity the ratio would be one and the circle would pass through the origin,  while an over-coupled cavity would have a ratio greater than one the origin would lie inside the circle. The point on the circle that is closest to the origin shows when the cavity resonance condition is met. The FSR of the cavity can be measured by observing this point at higher order resonances of the fundamental mode. The blue semicircle in the diagram centered on the cavity resonance shows the frequency region of the complex reflectivity where the laser is within the cavity linewidth. The value of $\Delta\nu$ at each of the end points of this semicircle correspond to the half-width-half-maximum (HWHM) of the cavity resonance and can be used with the FSR measurement to calculate the total losses in the cavity. Therefore, by measuring an arc of the circle traced by scanning the frequency of the input field relative to the cavity resonance, $A$, $T_1$, and $f_0$ can all be measured, in situ, for a given Eigenmode position on the cavity mirror.

\subsection{Heterodyne Transmission Measurement}\label{sec:technique-theory}

Measuring the complex reflectivity is complicated by the fact that conventional photo-detectors only measure power, and not the phase of the electric field. Furthermore, the spatial coupling of the input field to the cavity must also be considered. With a cavity that is near critical coupling, even a small mismatch in the spatial mode of the incident laser with respect to the cavity can have a large effect on the measured power in reflection near resonance. 

We circumvent these issues by using heterodyne interferometry with an auxiliary local oscillator (LO) laser coupled into the cavity from the opposite side of the cavity. By interfering a probe field in reflection with the LO in transmission at the cavity mirror, only the portion of the probe field's light which is in the spatial mode of the cavity contributes to the beat-note amplitude and phase. To show this, we can express the mode of probe field in reflection of the cavity  $\ket{\Psi_{\rm Pr}}$, as a superposition of the fundamental cavity mode $\ket{\psi_{00}} $ and the higher order modes $\ket{\psi_{\rm HOM}} $.

\begin{equation}
    \ket{\Psi_{\rm Pr}}  = \sqrt{P_{\rm Pr}}\exp \{i(\omega_1 t + \phi)\} (\eta \,\mathcal{R}(\Delta\nu)\ket{\psi_{00}} + \sqrt{1-\eta^2} \ket{\psi_{\rm HOM}})
\end{equation}
Here $P_{\rm Pr}$ is the power of the probe laser incident on the cavity and $\omega_1$ is its angular frequency given by $\omega_1=2\pi(\nu_1+\Delta \nu)$ where $\nu_1$ is the laser frequency corresponding to the closest cavity resonance. The spatial overlap between the probe laser and the fundamental Eigenmode of the cavity is given by $\eta$ with $\eta^2$ giving the fraction of laser power in the spatial mode of the cavity. $\mathcal{R}(\Delta\nu)$ is the cavity reflectivity function given by Equation~\ref{Eq:E_r}.

In this case the field of the LO laser can just be represented by the fundamental mode of the cavity.
\begin{equation}
    \ket{\Psi_{\rm LO}}  = \sqrt{P_{\rm LO}}\exp \{i\omega_2 t\} \ket{\psi_{00}} 
\end{equation}
$P_{\rm LO}$ is then the LO power in transmission of the cavity and $\omega_2$ is the angular frequency of the LO laser which can be expressed as $\omega_2=2\pi(\nu_1-nf_0)$ with $nf_0$ being some non-zero integer multiple of the free spectral range of the cavity. Therefore the frequency difference between the probe laser and the LO is ${\Delta\omega=2\pi(nf_0+\Delta\nu})$.

With this the field reflected by the cavity is given by the following equation.
\begin{multline}
        \ket{\Psi_{\rm Pr}} + \ket{\Psi_{\rm LO}}=  e^{i\omega_2t}\left(\sqrt{P_{\rm Pr}}\eta e^{i(\Delta\omega t +\phi)}\,\mathcal{R}+\sqrt{P_{\rm LO}}\right)\ket{\psi_{00}} \\  + e^{i\omega_1 t}\sqrt{P_{\rm Pr}}\sqrt{1-\eta^2} \ket{\psi_{\rm HOM}}
\end{multline}
The power in reflection $P_{\rm R}$ of the cavity can then be found by calculating the absolute square of the field, keeping in mind that $\ket{\psi_{00}}$ and $\ket{\psi_{\rm HOM}}$ are orthogonal and their cross terms $\braket{\psi_{00}|\psi_{\rm HOM}} = \braket{\psi_{\rm HOM}|\psi_{00}}=0$, while $\braket{\psi_{00}|\psi_{00}} = \braket{\psi_{\rm HOM}|\psi_{\rm HOM}}=1$ .
\begin{multline}
  P_{\rm R} =
      2\sqrt{P_{\rm LO} P_{\rm Pr}}\eta\left\{{\rm Re}[\mathcal{R}]\cos(\Delta\omega t-\phi) + {\rm Im}[\mathcal{R}]\sin(\Delta\omega t-\phi)\right\}\\
    + \left(1 - \eta^2 \left(1-|\mathcal{R}|^2\right)\right) P_{\rm Pr}  + P_{\rm LO} 
\end{multline}
Here the terms on the first line show the interference beat-note between the two lasers oscillating at the difference frequency $\Delta\omega$, while the terms on the bottom line give the static reflected power. By measuring of the reflected power and then demodulating at $\Delta\omega$ with both sine and cosine functions, the real and imaginary parts of $\mathcal{R}$ can be measured as the probe field frequency is scanned over resonance.

We will refer to the data demodulated with the cosine function as the `in-phase' component of the power $P_I$, while the data demodulated with the sine function will be called the `quadrature' component $P_Q$. As the following equations show, the in-phase component of the power is dependent on the real part of $\mathcal{R}$ while the quadrature component gives the imaginary part of $\mathcal{R}$.
\begin{equation}
    P_{I} = \langle P_{\rm R} \cos (\Delta\omega)\rangle =  \eta \sqrt{P_{\rm LO} P_{\rm Pr}}~  \mathrm{Re}[e^{i\phi}\,\mathcal{R}]
\end{equation}
\begin{equation}
    P_{Q} = \langle P_{\rm R} \sin (\Delta\omega t)\rangle =  \eta \sqrt{P_{\rm LO} P_{\rm Pr}}\, \mathrm{Im}[e^{i\phi}\,\mathcal{R}]
\end{equation}
$P_I$ and $P_Q$ can be measured as the frequency $\Delta\omega$ is scanned over the resonance of the cavity. This typically is represented as an amplitude and phase response on a bode plot, where the amplitude response is given by $\sqrt{P_I^2 + P_Q^2}$ and the phase response is $\arctan\left(\frac{P_Q}{P_I}\right)$. The following equations show how we can derive the functional form of $\mathcal{R}$ from this. 
\begin{equation} \label{eq:abs R}
   \sqrt{P_I^2 + P_Q^2} = \eta \sqrt{P_{\rm LO} P_{\rm Pr}} \, |\mathcal{R}| 
\end{equation}
\begin{equation}
    \arctan\left(\frac{P_Q}{P_I}\right) =  \arg(e^{i\phi}\,\mathcal{R})
\end{equation}
While the term ${\eta \sqrt{P_{\rm LO} P_{\rm Pr}}}$ of Equation~\ref{eq:abs R} scales the measured amplitude, a fit can be performed over $\mathcal{R}$ on the complex plane with the scaling left as a free parameter. Likewise, the phase offset between the interference beat-note phase and the phase of the demodulation signal $\phi$ may also be left as a free parameter in this fit. Since the term inside the $\arctan$ is a power ratio, its clear that calculating  the amplitude and phase of $\mathcal{R}$ for a measurement requires no knowledge the calibration of the photodetectors measuring the beat-note, the power levels of the lasers, nor their spatial overlap to the cavity. 

Nevertheless, for poor spatial coupling of the lasers to the cavity, the shot noise contribution of the HOM will lead to a lower signal to noise ratio for the laser PDH and PLL control loops and the technique will not be shot noise limited. Therefore, care should be taken to ensure the highest spatial coupling possible when using this method for cavity characterization.

\begin{figure}
    \centering
    \includegraphics[width=\linewidth]{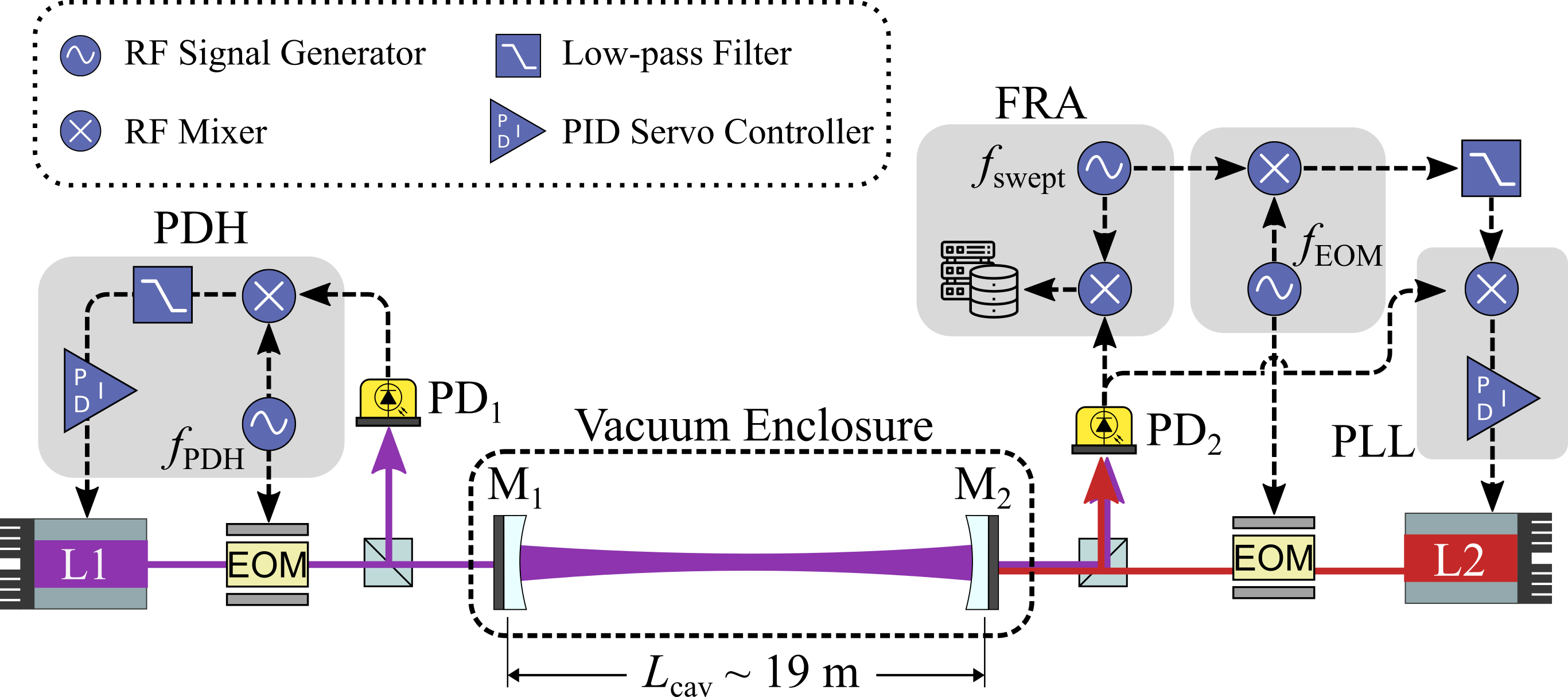}
    \caption{Schematic layout of the experiment, representing the configuration where Laser 1 serves as the resonant local oscillator and Laser 2 generates the probe field. This arrangement yields the cavity complex reflectivity from the side of Laser 2, and therefore $T_2$. To measure the complex reflectivity from the side of Laser 1, and thus $T_1$, the control loops which provide signals to the EOMs and lasers are exchanged.}
    \label{fig:experimental_layout}
\end{figure}

\section{Methods}

A diagram of the optical system, control electronics, and readout scheme for the experiment is shown in Figure \ref{fig:experimental_layout}. The cavity used in this study was formed by two mirrors mounted on separate optical benches within a shared vacuum enclosure consisting of two end tanks connected by a beam tube. 

Two lasers, Laser 1 and Laser 2, are located on opposite sides of the cavity, on the same optical tables as Mirror 1 and Mirror 2, respectively. Laser 1 is non-planar ring oscillator (NPRO) while Laser 2 is a master oscillator power amplifier (MOPA) seeded by a second NPRO. Both are operated at a wavelength of 1064\,nm. For the purpose of this measurement,  $\sim$10\,mW from each laser is injected to the cavity. Each laser passes through a resonant electro-optic modulator (EOM) to generate phase-modulation sidebands prior to being coupled into the cavity.

\subsection{Cavity Design}
The cavity mirrors, acquired from the company \textit{LaserOptik GmbH}, are nominally identical, each with a diameter of 50.8 mm. Their coatings consist of alternating silica/tantala dielectric layers deposited on a polished fused silica substrate. The cavity is assembled in a nearly confocal geometry. A free spectral range of $7.89160\pm0.00001$\,MHz was measured, giving a single-pass optical path length of $18.99440\pm0.00002$\,m. The radii of curvature of the cavity mirrors were obtained by measuring the frequency spacing between the cavity's fundamental and first-order transverse modes. This indicated that the average of the radii of curvature of the two mirrors is 19.95\,m. The mirror mounts for the cavity optics could be controlled remotely via piezo actuators, allowing for precise adjustments of the alignment of each mirror while under vacuum.

\begin{figure}
    \centering
    \includegraphics[width=0.4\linewidth]{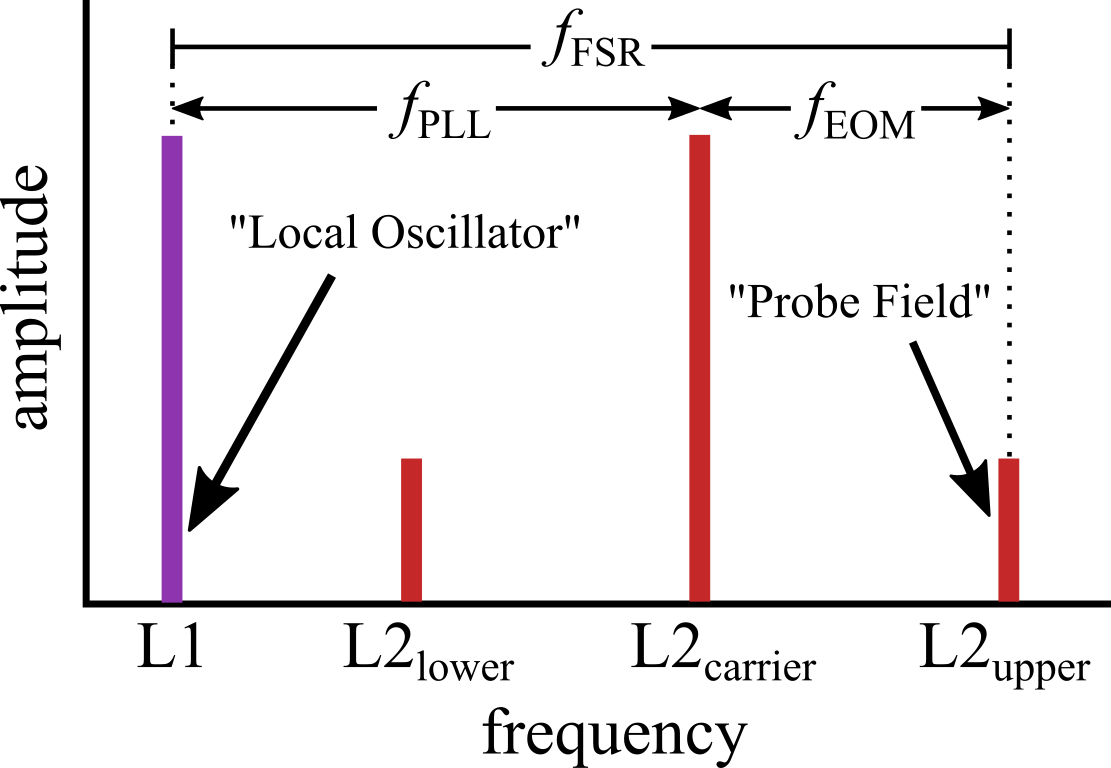}
    \caption{Diagram representing the relative frequencies of the individual fields in reflection of M$_2$. The local oscillator field is provided by Laser 1 in transmission of the cavity while the upper sideband of Laser 2, located approximately one FSR away, is used as the probe field. Using the PLL to maintain the relative frequency between L1 and the carrier of L2 $f_\mathrm{PLL}$, the amplitude and phase of the beat-note between the local oscillator and the swept probe field is measured at each frequency point of the scan.}
    \label{fig:F_s}
\end{figure}

\subsection{Measurement Technique}

In order to perform the heterodyne cavity reflectivity measurement described in Section \ref{sec:technique-theory}, the frequency of Laser 1 was controlled with respect to the resonance of the cavity utilizing the Pound-Drever-Hall technique \cite{black2001}. The field from Laser 1 in transmission of the cavity serves as our local oscillator. In order to control the frequency of the probe field and suppress environmental disturbances, the Laser 2 carrier frequency is stabilized relative to the cavity-transmitted field of Laser 1 using an offset phase-locked loop (PLL) \cite{pll1999}. A resonant EOM is then used to generate phase-modulation sidebands on Laser 2 with modulation frequency $f_\mathrm{EOM}$. The upper sideband is used as the probe field which is scanned in frequency over a fundamental cavity resonance. For reference Figure~\ref{fig:F_s} shows the how the beat-note frequencies are configured in the setup.

The complex reflectivity of the cavity $\mathcal{R}$ is encoded in the amplitude and phase of the beat-note generated between the local oscillator and the scanned probe field.  This is measured using a digital Frequency Response Analyzer (FRA) that generates a sine wave with frequency $f_\mathrm{swept}$. This frequency is then swept over the cavity's first free spectral range $f_0 \approx 7.892$ MHz and the beat-note amplitude and phase are measured at points near the cavity resonance. To optimize the measurement resolution, we set the span of the sweep to be roughly a cavity linewidth of approximately $200 \; \rm Hz$.

The PLL offset frequency is generated by digitally mixing the FRA output swept sine with the EOM modulation frequency and low-pass filtering, resulting in a signal with frequency $f_\mathrm{PLL} = f_\mathrm{swept} - f_\mathrm{EOM}$. Therefore as $f_\mathrm{swept}$ is scanned, the offset between Laser 1 and Laser 2 changes accordingly.

The frequency control scheme described here allows the Laser 2 carrier frequency to remain off resonance of the cavity as the measurement is performed. This is preferred over a simplified method wherein Laser 2 is directly locked to Laser 1 in transmission and scanned over resonance for two reasons: for a nearly-impedance matched cavity, the amplitude of the beat-note used to control the frequency of Laser 2 diminishes, causing the PLL to become unstable once the probe field approaches resonance; secondly, the phase information of the beat-note is obfuscated by the PLL, which actively works to correct for any relative deviations in phase between the two fields. 

The beat-note amplitude and phase between Laser 1 and the probe field are then measured as a function of frequency in the FRA by demodulating the signal with $f_\mathrm{swept}$ as it is scanned over the cavity resonance. In order to measure the cavity reflectivity from M$_1$, the roles of Laser 1 and Laser 2 (and their associated control loops) are exchanged and the measurement is repeated. 

\begin{figure}[t]
        \centering
        \includegraphics[width=0.75\linewidth]{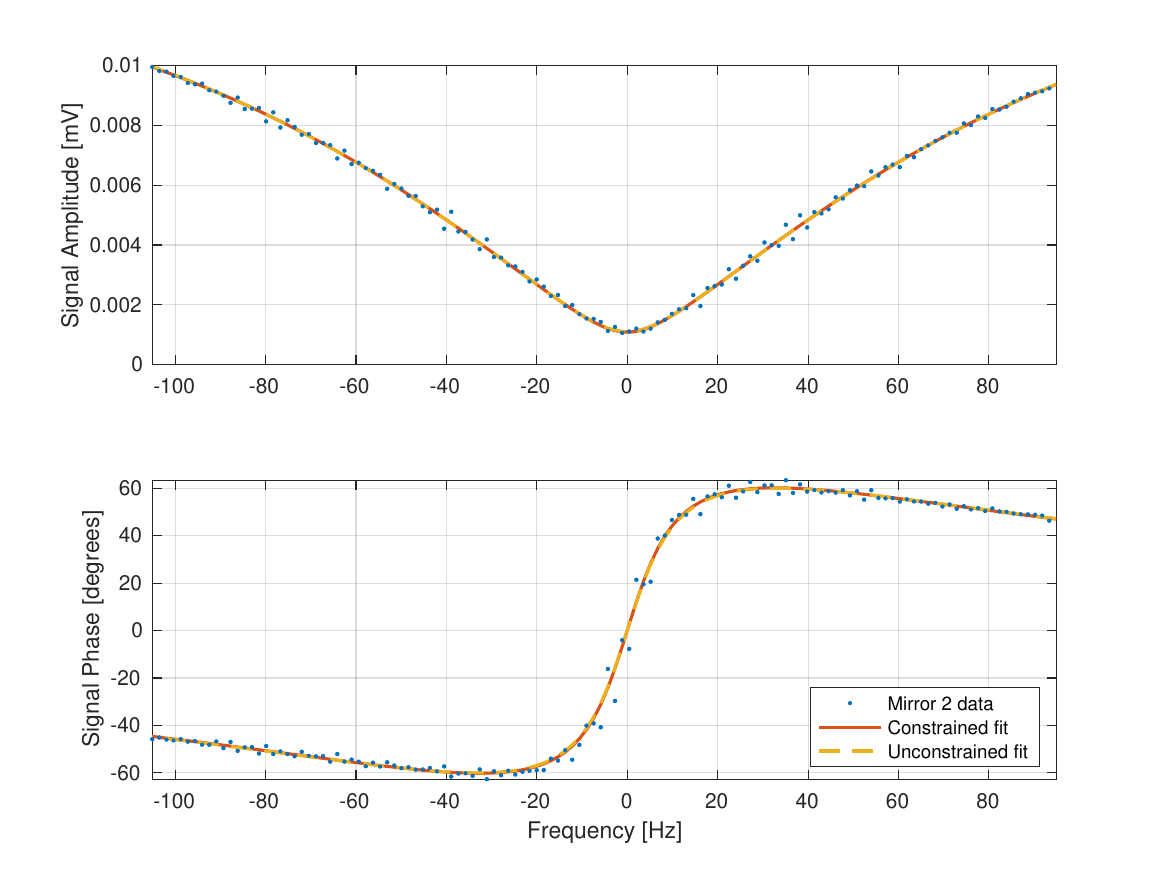}
    \caption{A single measurement of the amplitude (top) and phase (bottom) of the signal when scanning over the cavity resonance in reflection of mirror 2 is shown as the blue points. Due to the nature of this measurement the signal amplitude is proportional to the field strength in reflection of the cavity. Two different fitting methods, which are discussed in the following section, are applied to the data. A fit in which the storage time of the cavity is constrained to a value obtained using another method is shown as the orange line. The yellow line shows a fit in which the storage time was left as a free parameter. As the plot shows, the two methods agree very well.}
    \label{Fig:A Phi}
\end{figure}

\section{Results}\label{analysis}

The results of a single frequency response measurement are given in Figure~\ref{Fig:A Phi}. The upper plot gives the amplitude of the interference beat-note as it is scanned over the resonance of the cavity, while the lower plot is of the phase. 

In order to calculate the cavity parameters $T_i$, $A$, and $f_0$ from this measurement, the data was expressed as complex numbers and a function in the form of $\mathcal{R}(\Delta\nu)$ (Equation \ref{Eq:E_r}) was fit to it. Two separate fitting routines were performed to compare their results. The first, referred to as the `unconstrained fit,' allowed the transmissivity of the input mirror $T$, the round trip losses $A$, the FSR of the cavity, and the offset phase and amplitude of the input field $E_i$ to be free parameters. The second method, which we call the `constrained fit,' constrained the value of the total round trip losses $A$ based on the storage time measured directly before and after the measurement of $\mathcal{R}(\Delta\nu)$. The storage time measurements were performed by disengaging the frequency stabilization system of the local oscillator laser and observing the exponential decay of the power in transmission of the cavity.

The results of these fits of the measured data to  $\mathcal{R}(\Delta\nu)$ are shown in Figure~\ref{Fig:A Phi} with the unconstrained fit represented by the red trace and the constrained fit as the yellow dashed trace. In both cases, the two fitting methods agree well enough that it is difficult to visually distinguish them on these plots.

\begin{figure}
    \centering
    \begin{subfigure}[t]{0.49\textwidth}
        \centering
        \includegraphics[width=1\textwidth]{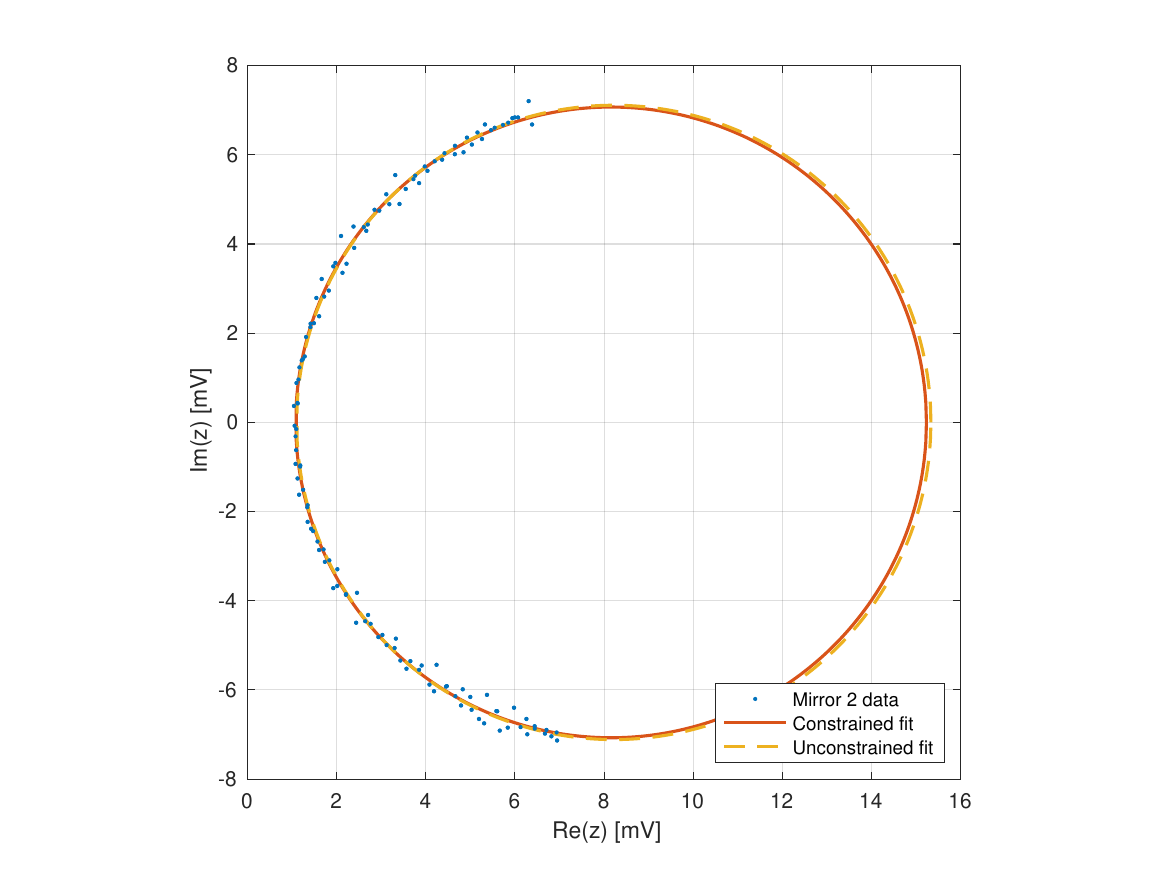}
        \caption{Two dimensional representation }
        \label{Fig:2D}
    \end{subfigure}%
    ~ 
    \begin{subfigure}[t]{0.5\textwidth}
        \centering
        \includegraphics[width=1\textwidth]{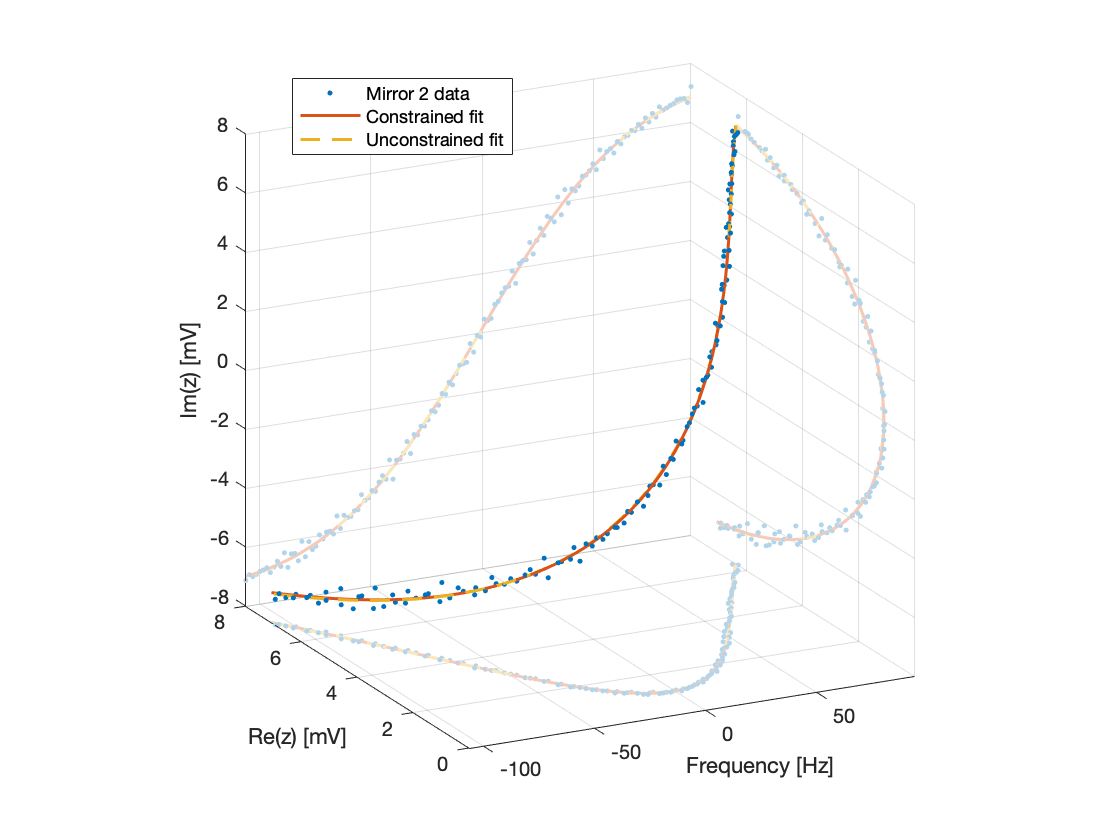}
        \caption{Three dimensional representation }
        \label{Fig:3D}
    \end{subfigure}
    \caption{A single measurement of the complex Lorentzian in reflection of mirror 2 when scanning over the cavity resonance is shown on the complex plane (a), as well as in three dimensions (b) where the additional axis represents the frequency of the probe field with respect to the cavity resonance. In both of these plots the blue points represent the measured data, while the orange line gives the resulting fit in which the total round-trip cavity losses are constrained to a value obtained by a storage time measurement, while the yellow line is a fit in which the round-trip losses where left as a free parameter.}
\end{figure}

\subsection{Data Analysis}

The fit was performed by calculating the minimum of the following expression with the Matlab function $\mathtt{lsqnonlin}$. 
\begin{equation}
    \left|X_1 +iX_2-\left(a_1+ia_2\right)\left(1-\frac{a_3}{a_4+i\frac{2\pi\left(X_0 - f_0 - a_0  \right)}{f_0}}\right)\right|
    \label{Eq:fit}
\end{equation}
Here $X_1$ and $X_2$ represent the in-phase and quadrature components of the measured data, while $X_0$ is the difference frequency between the probe field and the LO. The variables denoted by $a$ represent the fitting parameters. In this case $a_1+ia_2$ gives the location of the interference beat-note on the complex plane, $a_3$ is analogous to $T_i$, $a_4$ is $A/2$, and $a_0+f_0$ gives the precise FSR of the cavity. Here $f_0$ is a fixed parameter representing the initial measurement of the cavity FSR which is accurate to 100 Hz. 

Properly estimating the initial parameters is critical to ensuring that the fit finds the minimum of Equation~\ref{Eq:fit}. To do this the initial values for the $a$ parameters of the fit were estimated by calculating the closest arc on the complex plane resembling the data independent of the frequency information at each point. 

From this the initial estimate of the parameters $a_1$, $a_2$, and $a_3$ could be provided. The measurement of the FSR was used for $f_0$ with the initial value for $a_0$ being 0. In the case of the constrained fit $a_4$ was constrained based on the storage time measurement, while for the unconstrained fit the storage time measurement was used to estimate its initial value

To efficiently find the minimum value of equation \ref{Eq:fit}, not all parameters were fit at once. Rather, for the unconstrained fit first a fit was performed with $a_4$ and $a_0$ as free parameters, with only $a_0$ as the free parameter for the constrained fit, while $a_1$, $a_2$, and $a_3$ were fixed at their initial estimated values. Following this, the parameters $a_0$ and $a_4$ where fixed at the values obtained by the fit and  $a_1$, $a_2$, and $a_3$ were free fitting parameters.


Figure\,\ref{Fig:3D} shows the data measured on the complex plane as blue points much like the diagram in Figure\,\ref{fig:circle-cartoon}. The red curve shows the full circle resulting from the constrained fit, while the yellow circle is the result of the unconstrained fit. The agreement between the data and the fits are apparent from this plot.

This same data is plotted in three dimensions in Figure \ref{Fig:3D} with the frequency dependence represented by the third axis. The results of the constrained and unconstrained fits are shown again as the red and yellow curves, and again the fits agree well with the measured data. The three panels in the background of this figure show the relation between the three dimensional representation of this plot, the two dimensional arc of Figure~\ref{Fig:2D}, and the Bode plot in Figure~\ref{Fig:A Phi}. The upper right panel of Figure~\ref{Fig:3D} shows how the data represented independent of the frequency information gives the two dimensional arc of $\mathcal{R}$. The bottom panel shows how, when the frequency is close to the resonance of the cavity, the plot of the real component of $\mathcal{R}$ versus the frequency approximates to the amplitude data of the Bode plot in Figure~\ref{Fig:A Phi}. The upper left panel of Figure~\ref{Fig:3D} then shows how the plot of the imaginary component of $\mathcal{R}$ versus the frequency approximates to the phase data of the Bode plot near the cavity resonance.

\subsection{Measured Cavity Parameters}

\begin{table}[]
    \footnotesize
    \centering
    \begin{tabular}{| c c c | c c c |} 
        \multicolumn{3}{c}{Mirror 1} & \multicolumn{3}{c}{Mirror 2}    \\  \hline
        parameter   &method         & value (ppm) &               
        parameter   &method         & value (ppm)      \\    \hline
        $A$       &storage time   & $195.2\pm0.5$    &
        $A$       &storage time   & $194.9\pm0.4$    \\
        $A$       &unconstrained   & $197.5\pm0.5$    &
        $A$       &unconstrained   & $193.0\pm0.4$    \\
        $T_1$     &constrained & $89.5\pm0.2$      &
        $T_2$     &constrained & $90.2\pm0.2$     \\ 
        $T_1$     &unconstrained & $90.5\pm0.2$     &
        $T_2$     &unconstrained &  $89.4\pm0.2$    \\
    \hline
    \end{tabular}
    
    \caption{The resulting total round-trip losses ($A$), and mirror transmissivities ($T_1$ and $T_2$) obtained from applying the two fitting methods to the measurements of the cavity mirrors.     
    }
        \label{tab:cav}
\end{table}

Table~\ref{tab:cav} shows the results for the total losses $A$ and input mirror transmissivities $T_1$ and $T_2$, measured using the methods in the previous section. Here the first value for the total losses in each mirror is calculated from the storage time measurement made before and after the frequency scans were performed. The storage times of $1.298\pm0.003$\,ms and $1.300\pm0.002$\,ms correspond to total losses of $195.2\pm0.5$\,ppm and $194.9\pm0.4$\,ppm obtained at the time of the measurements of Mirror 1 and 2 respectively. In this case, we believe that the drift of the eigenmode position over the course of the measurement and in the time between measuring Mirror 1 and Mirror 2 is responsible for the error, as well as slight change of the storage time. 

The unconstrained fits for the frequency scan measured total losses of $197.5\pm0.5$\,ppm for Mirror 1 and $193.0\pm0.4$\,ppm for Mirror 2. The error bars for these measurements was calculated based on the statistical uncertainty of 41 consecutive measurements for Mirror 1 and 50 measurements for Mirror 2. While the results of the unconstrained fit agree with the storage time measurement to an accuracy of nearly 1\% in both mirrors, it is also apparent that the error bars of these two measurements do not overlap. This discrepancy is not completely understood.

For the measurement of the transmissivity of Mirror 1, the constrained fit gave $89.5\pm0.2$\,ppm, while the unconstrained fit gave $90.5\pm0.2$\,ppm. For Mirror 2 the constrained fit found a transmissivity of $90.2\pm0.1$\,ppm, with the unconstrained fit giving $89.4\pm0.2$\,ppm.  These results show that for both mirrors the constrained and unconstrained fits gave a transmissivity to within 1\,ppm of each other. 
In the measurements of both mirrors the error bars on the constrained fits shown in the table were determined by the uncertainty on the total losses from the storage time measurements. This error was significantly higher than the statistical uncertainty over series of frequency scans of each mirror. In the case of the unconstrained fits the error of these measurement has no dependence on the uncertainty in the storage time and is rather determined by the statistical  error in the frequency scan measurements.

It should also be noted that if a selection criteria was applied to the unconstrained fits in which only measurements with a $\chi^2$ less than the mean $\chi^2$ for all measurements for a mirror particular were used, the values for $A$ and $T$ were closer to the results obtained by the storage time measurement and constrained fit, than when no selection criteria was applied. For Mirror 1 with this selection criteria the unconstrained fit gave $A=197.1\pm0.5$ and $T_1=90.2\pm0.3$. For Mirror 2, applying the selection criteria lead to $A=194.0\pm0.5$ and $T_1=89.8\pm0.2$. Applying the selection criteria to the constrained fits had no significant impact on their results. 

It is suspected that due to the simplicity of measuring the total losses via the storage time technique, this method when combined with the constrained fit of the frequency scan, is the more robust way of identifying these parameters. The fact that the selection criteria had no impact on the result with this technique, while it did push the results using the constrained fit closer to the values obtained with the storage time and constrained fit appears to reinforce this conclusion.

Furthermore, one of the primary sources of noise in these measurements is thought to be the length changes of the cavity itself during the measurement. Over the 30\,s measurement time these length changes could lead to changes in the FSR, thus detuning the setup and distorting the shape of the resonance. 

We would also like to mention that the combined transmissivities of the the mirrors of the cavity reported here of $180\pm1$\,ppm with corresponding total losses of $195\pm2$\,ppm give excess losses in addition to the mirror transmissivities of $15\pm2$\,ppm. With a length of 19.007\,m this is better than 1\,ppm/m excess loss per unit length, which is on par with the lowest loss cavities in the world for such lengths \cite{isogai13}.

\section{Conclusion}

Here we have shown how the cavity free spectral range, total-round trip losses, and individual mirror transmissivities can all be derived from a measurement of the complex reflectivity using a transmitted local oscillator field. These parameters can be measured in-situ at the spatial Eigenmode position of operation, are independent of the spatial coupling of the lasers to the cavity, and can be used with any cavity impedance matching configuration. This technique is demonstrated on a 19\,m, high-finesse optical resonator and is capable of measuring the cavity mirror losses and transmissivities with an accuracy on the order of several ppm. The total round-trip attenuation agrees with measurements made using cavity ring-downs to 1\%.

This technique is widely applicable to optical cavities of all lengths. While it is shown here on a resonator with a relatively long storage time of 1.30\,ms, it may be even more accurate for shorter storage time cavities due to the reduced length noise over the measurement time. With respect to longer baseline experiments, the next generation of gravitational wave detectors could also be well suited to utilize this technique to initially characterize their core optics. Finally, this method could prove to be important for light-shining-through-a-wall style experiments such as ALPS\,II, where the calibration of the experimental sensitivity relies on the precise knowledge of the cavity mirror transmissivites.

\subsection*{Funding}
The work is supported by the Deutsche Forschungsgemeinschaft (DFG, German Research Foundation) under Germany’s Excellence Strategy – EXC 2121 ``Quantum Universe" – 390833306 and - grant number WI 1643/2-1, the Partnership for Innovation, Education and Research (PIER) of DESY and Universit\"at Hamburg under PIER Seed Project - PIF-2022-18, the German Volkswagen Stiftung, the National Science Foundation - grant numbers PHY-2110705 365 and PHY-1802006, the Heising Simons Foundation - grant numbers 2015-154 and 2020-1841, the UK Science and Technologies Facilities Council - grant number ST/T006331/1.

\subsection*{Acknowledgments}
The authors would like to thank the members of the ALPS Collaboration and especially Benno Willke and Guido Mueller for many illuminating conversations.



\printbibliography 

\end{document}